\documentclass{article}

\usepackage{microtype}
\usepackage{graphicx}
\usepackage{subcaption}
\usepackage{booktabs} 

\usepackage{hyperref}

\usepackage[english]{babel}
\usepackage{blindtext}
\usepackage{amsmath}
\usepackage{amssymb}
\usepackage{bm}
\usepackage{xcolor}
\usepackage{multirow}
\usepackage{arydshln}

\renewcommand{\vec}[1]{\mathbf{#1}}
\newcommand{\boldpar}[1]{\textbf{#1.}}

\newcommand{\ULsubfloat}[2][\empty]
{\hbox{
  \sbox0{#2}
  \captionsetup{font=scriptsize, position=top, justification=centering, singlelinecheck=false, labelformat=empty}%
  \rotatebox[origin=bl]{90}{\begin{minipage}[b]{\dimexpr \ht0+\dp0}
    \subcaption{#1}
  \end{minipage}}\raisebox{\dp0}{\usebox0}%
}}

\usepackage[accepted]{icml2020}

\icmltitlerunning{PolyGen: An Autoregressive Generative Model of 3D Meshes}

\begin{document}

\twocolumn[
\icmltitle{PolyGen: An Autoregressive Generative Model of 3D Meshes}



\icmlsetsymbol{equal}{*}

\begin{icmlauthorlist}
\icmlauthor{Charlie Nash}{dm}
\icmlauthor{Yaroslav Ganin}{dm}
\icmlauthor{S. M. Ali Eslami}{dm}
\icmlauthor{Peter W. Battaglia}{dm}
\end{icmlauthorlist}

\icmlaffiliation{dm}{DeepMind, London, United Kingdom}

\icmlcorrespondingauthor{Charlie Nash}{charlienash@google.com}

\icmlkeywords{Machine Learning, ICML}
\vskip 0.3in
]



\printAffiliationsAndNotice{}  

\begin{abstract}
Polygon meshes are an efficient representation of 3D geometry, and are of central importance in computer graphics, robotics and games development. Existing learning-based approaches have avoided the challenges of working with 3D meshes, instead using alternative object representations that are more compatible with neural architectures and training approaches. We present an approach which models the mesh directly, predicting mesh vertices and faces sequentially using a Transformer-based architecture. Our model can condition on a range of inputs, including object classes, voxels, and images, and because the model is probabilistic it can produce samples that capture uncertainty in ambiguous scenarios. We show that the model is capable of producing high-quality, usable meshes, and establish log-likelihood benchmarks for the mesh-modelling task. We also evaluate the conditional models on surface reconstruction metrics against alternative methods, and demonstrate competitive performance despite not training directly on this task.

\end{abstract}

\section{Introduction}
\label{introduction}
Polygon meshes are an efficient representation of 3D geometry, and are widely used in computer graphics to represent virtual objects and scenes. Automatic mesh generation enables more rapid creation of the 3D objects that populate virtual worlds in games, film, and virtual reality. In addition, meshes are a useful output in computer vision and robotics, enabling planning and interaction in 3D space. 

Existing approaches to 3D object synthesis rely on the recombination and deformation of template models \cite{DBLP:journals/tog/KalogerakisCKK12, DBLP:journals/tog/ChaudhuriKGK11}, or a parametric shape family \cite{DBLP:journals/cgf/SmelikTBB14}. Meshes are challenging for deep learning architectures to work with because of their unordered elements and discrete face structures. Instead, recent deep-learning approaches have generated 3D objects using alternative representations of object shape---voxels \cite{DBLP:conf/eccv/ChoyXGCS16}, pointclouds, occupancy functions \cite{DBLP:conf/cvpr/MeschederONNG19}, and surfaces \cite{DBLP:conf/cvpr/GroueixFKRA18}---however mesh reconstruction is left as a post-processing step and can yield results of varying quality. This contrasts with the human approach to mesh creation, where the mesh itself is the central object, and is created directly with 3D modelling software. Human created meshes are compact, and reuse geometric primitives to efficiently represent real-world objects. 
\begin{figure}[t]
    \centering
    \includegraphics[width=\linewidth]{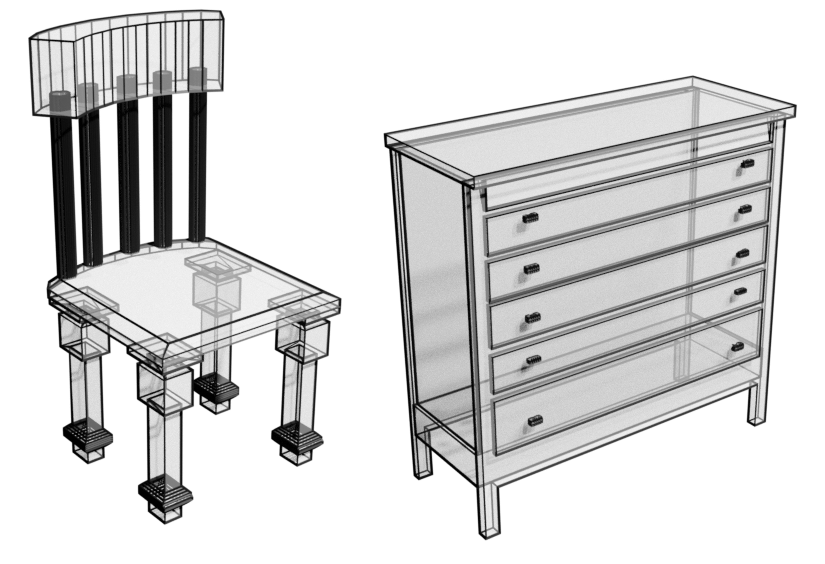}
    \caption{Class conditional $ n $-gon meshes generated by PolyGen.}
    \label{fig:teaser}
\end{figure}
\begin{figure*}[ht]
    \centering
    \includegraphics[width=\linewidth]{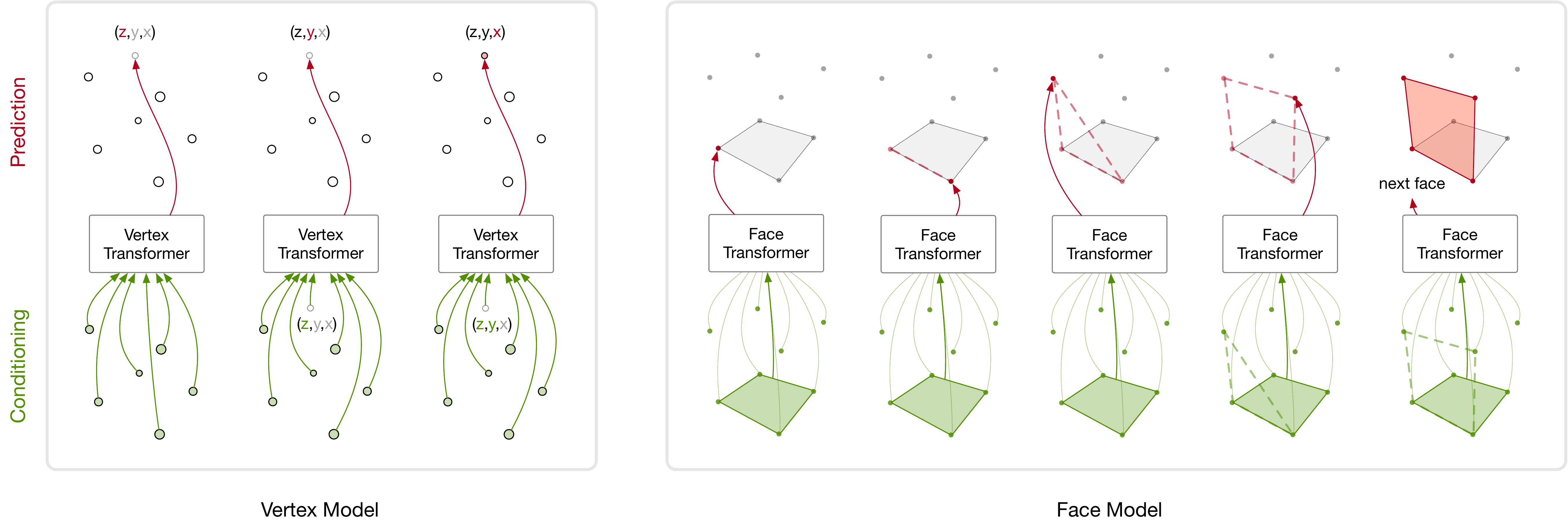}
    \caption{PolyGen first generates mesh vertices (left), and then generates mesh faces conditioned on those vertices (right). Vertices are generated sequentially from lowest to highest on the vertical axis. To generate the next vertex the current sequence of vertex coordinates is passed as context to a vertex Transformer, which outputs a predictive distribution for the next vertex coordinate. The face model takes as input a collection of vertices, and the current sequence of face indices, and outputs a distribution over vertex indices.}
    \label{fig:polygen}
\end{figure*}
Neural autoregressive models have demonstrated a remarkable capacity to model complex, high-dimensional data including images \cite{DBLP:conf/icml/OordKK16}, text \cite{radford2019language} and raw audio waveforms \cite{DBLP:conf/ssw/OordDZSVGKSK16}. Inspired by these methods we present PolyGen, a neural generative model of meshes, that autoregressively estimates a joint distribution over mesh vertices and faces.

PolyGen consists of two parts: A vertex model, that unconditionally models mesh vertices, and a face model, that models the mesh faces conditioned on input vertices. Both components make use of the Transformer architecture \cite{DBLP:conf/nips/VaswaniSPUJGKP17}, which is effective at capturing the long-range dependencies present in mesh data. The vertex model uses a masked Transformer decoder to express a distribution over the vertex sequences. For the face model we combine Transformers with pointer networks \cite{DBLP:conf/nips/VinyalsFJ15} to express a distribution over variable length vertex sequences. 

We evaluate the modelling capacity of PolyGen using log-likelihood and predictive accuracy as metrics, and compare statistics of generated samples to real data. We demonstrate conditional mesh generation with object class, images and voxels as input and compare to existing mesh generation methods. Overall, we find that our model is capable of creating diverse and realistic geometry that is directly usable in graphics applications. 
\section{PolyGen}
\label{model}
Our goal is to estimate a distribution over meshes $\mathcal{M}$ from which we can generate new examples. A mesh is a collection of 3D vertices $\mathcal{V}$, and polygon faces $F$, that define the shape of a 3D object. We split the modelling task into two parts: i) Generating mesh vertices $\mathcal{V}$, and ii) generating mesh faces $\mathcal{F}$ given vertices. Using the chain rule we have:
\begin{align}
    p(\mathcal{M}) &= p(\mathcal{V}, \mathcal{F}) \\
    &= p(\mathcal{F} | \mathcal{V})p(\mathcal{V})
\end{align}
We use separate vertex and face models, both of which are autoregressive; factoring the joint distribution over vertices and faces into a product of conditional distributions. To generate a mesh we first sample the vertex model, and then pass the resulting vertices as input to the face model, from which we sample faces (see Figure~\ref{fig:polygen}). In addition, we optionally condition both the vertex and face models on a context $\vec{h}$, such as the mesh class identity, an input image, or a voxelized shape. 
\begin{figure}[t]
    \centering
    \begin{subfigure}[b]{0.475\linewidth}
        \centering
        \includegraphics[width=0.85\linewidth]{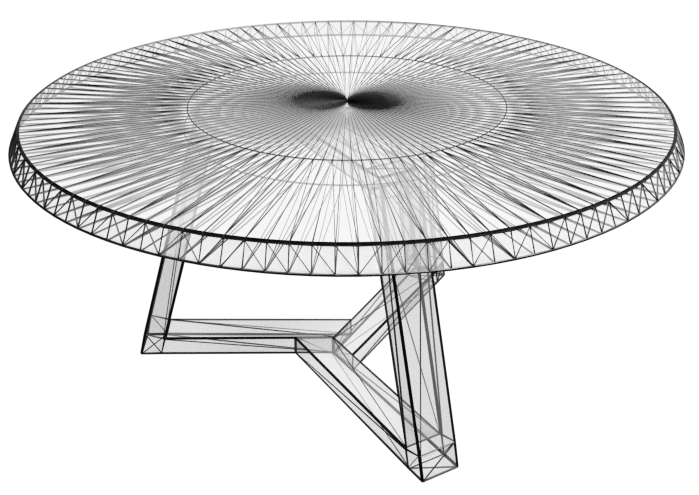}
        \caption{Triangle mesh}
    \end{subfigure}
    \begin{subfigure}[b]{0.475\linewidth}
        \centering
        \includegraphics[width=0.85\linewidth]{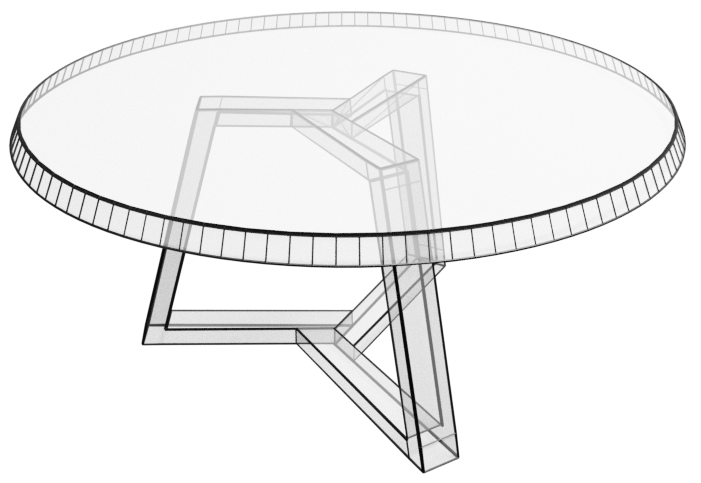}
        \caption{$n$-gon mesh}
    \end{subfigure}
    \caption{Triangle meshes consist entirely of triangles. $n$-gon meshes efficiently represent shapes using variable size polygons. 
    }
    \label{fig:ngon}
\end{figure}

\subsection{$n$-gon Meshes}
\label{subsec:mesh-representation}
3D meshes typically consist of collections of triangles, but many meshes can be more compactly represented using polygons of variable sizes. Meshes with variable length polygons are called $n$-gon meshes:  
\begin{align}
    \mathcal{F}_{\text{tri}} &= \left\{\left(f_1^{(i)}, f_2^{(i)}, f_3^{(i)}\right)\right\}_i \\
    \mathcal{F}_{n\text{-gon}} &= \left\{\left(f_1^{(i)}, f_2^{(i)}, \dotsc,  f_{N_i}^{(i)}\right)\right\}_i
\end{align}
where $N_i$ is the number of faces in the $i$-th polygon and can vary for different faces. This means that large flat surfaces can be represented with a single polygon e.g.\ the top of the circular table in Figure~\ref{fig:ngon}. In this work we opt to represent meshes using $n$-gons rather than triangles. This has two main advantages: The first is that it reduces the size of meshes, as flat surfaces can be specified with a reduced number of faces. Secondly, large polygons can be triangulated in many ways, and these triangulations can be inconsistent across examples. By modelling $n$-gons we factor out this triangulation variability. 

A caveat to this approach is that $n$-gons do not uniquely define a 3D surface when $n$ is greater than 3, unless the vertices it references are planar. When rendering non-planar $n$-gons, polygons are first triangulated by e.g.\ projecting vertices to a plane \cite{DBLP:journals/algorithmica/Held01}, which can cause artifacts if the polygon is highly non-planar. In practice we find that most of the $n$-gons produced by our model are either planar, or close to planar, such that this is a minor issue. Triangle meshes are a subset of $n$-gon meshes, and PolyGen can therefore be used to model them if required. 

\subsection{Vertex Model}
\label{subsec:vertex-model}
The goal of the vertex model is to express a distribution over sequences of vertices.  We order the vertices from lowest to highest by $z$-coordinate, where $z$ represents the vertical axis. If there are vertices with the same $z$-value, we order by $y$ and then by $x$ value. After re-ordering, we obtain a flattened sequence by concatenating tuples of $(z_i, y_i, x_i)_i$ coordinates. Meshes have variable numbers of vertices, so we use a stopping token $s$ to indicate the end of the vertex sequence. We denote the flattened vertex sequence $\mathcal{V}^{\textrm{seq}}$ and its elements as $v_n , n=1, \dotsc, N_V$. We decompose the joint distribution over $\mathcal{V}^{\textrm{seq}}$ as the product of a series of conditional vertex distributions:
\begin{align}
    p(\mathcal{V}^{\textrm{seq}} ; \theta) & = \prod_{n=1}^{N_V} p(v_n | v_{<n} ; \theta)
\end{align}
We model this distribution using an autoregressive network that outputs at each step the parameters of a predictive distribution for the next vertex coordinate. This predictive distribution is defined over the vertex coordinate values as well as over the stopping token $s$. The model is trained to maximize the log-probability of the observed data with respect to the model parameters $\theta$.

\boldpar{Architecture} The basis of the vertex model architecture is a Transformer decoder \cite{DBLP:conf/nips/VaswaniSPUJGKP17}, a simple and expressive model that has demonstrated significant modeling capacity in a range of domains \cite{DBLP:journals/corr/abs-1904-10509, DBLP:conf/iclr/HuangVUSHSDHDE19, DBLP:conf/icml/ParmarVUKSKT18}. Mesh vertices have strong non-local dependencies, with object symmetries and repeating parts, and the Transformer's ability to aggregate information from any part of the input enables it to capture these dependencies. We use the improved Transformer variant with layer normalization inside the residual path, as in \cite{DBLP:journals/corr/abs-1904-10509, DBLP:journals/corr/abs-1910-06764}. See Figure~\ref{fig:vertex-model} in the appendix for an illustration of the vertex model and appendix C for a full description of the Transformer blocks.
\begin{figure}[t]
    \centering
    \captionsetup[subfigure]{labelformat=empty, font=scriptsize}
    \ULsubfloat[Lamp]{%
       \includegraphics[width=0.92\linewidth]{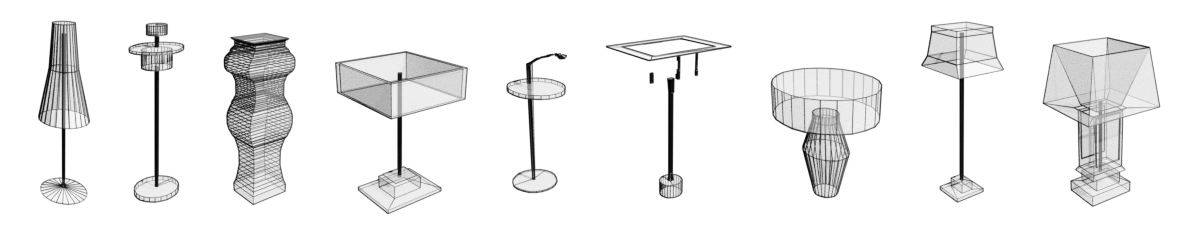}%
    }
    \ULsubfloat[Sofa]{%
       \includegraphics[width=0.92\linewidth]{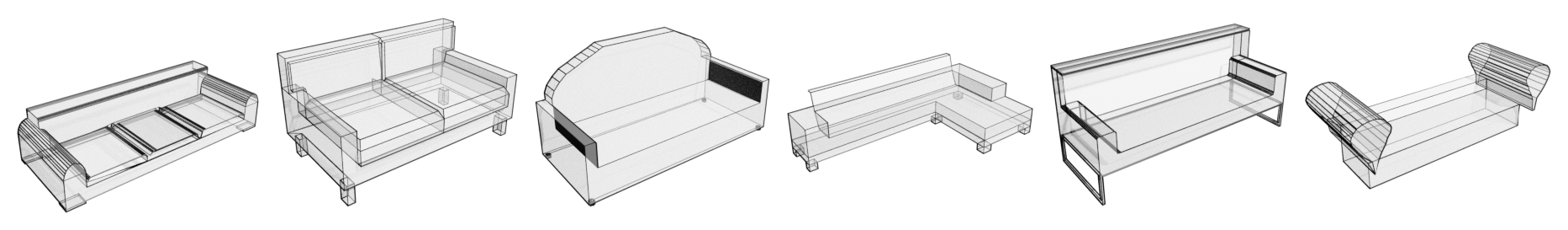}%
    }
    \ULsubfloat[Clock]{%
      \includegraphics[width=0.92\linewidth]{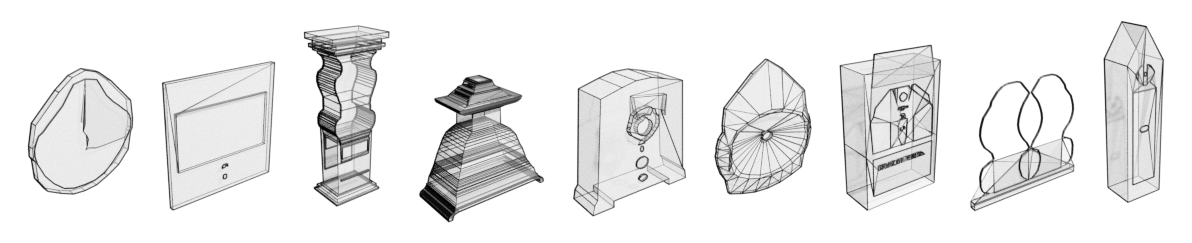}%
    }
    \ULsubfloat[Cabinet]{%
      \includegraphics[width=0.92\linewidth]{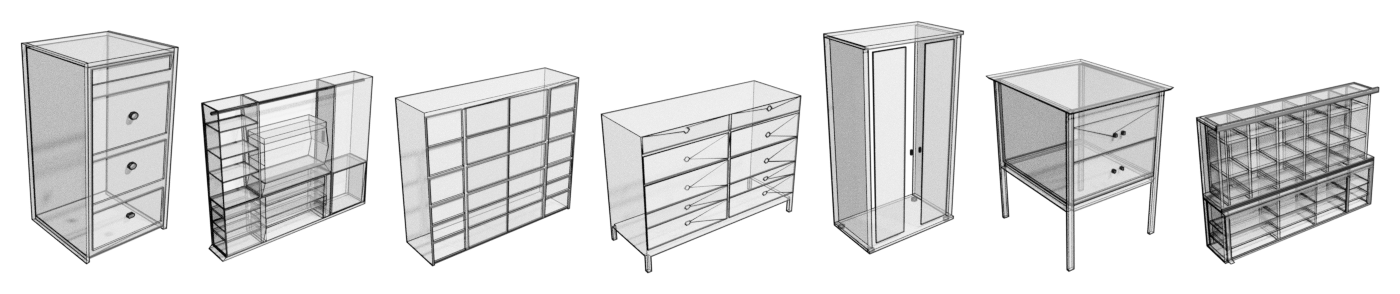}%
    }
    \ULsubfloat[Jar]{%
      \includegraphics[width=0.92\linewidth]{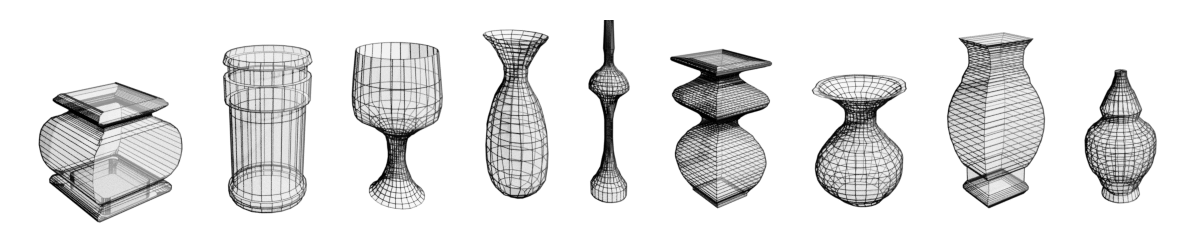}%
    }
    \ULsubfloat[Bench]{%
      \includegraphics[width=0.92\linewidth]{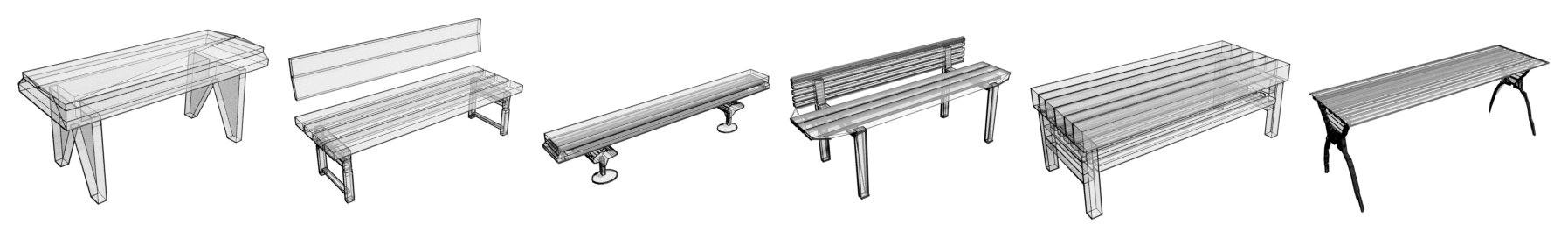}%
    }
    \ULsubfloat[Monitor]{%
      \includegraphics[width=0.92\linewidth]{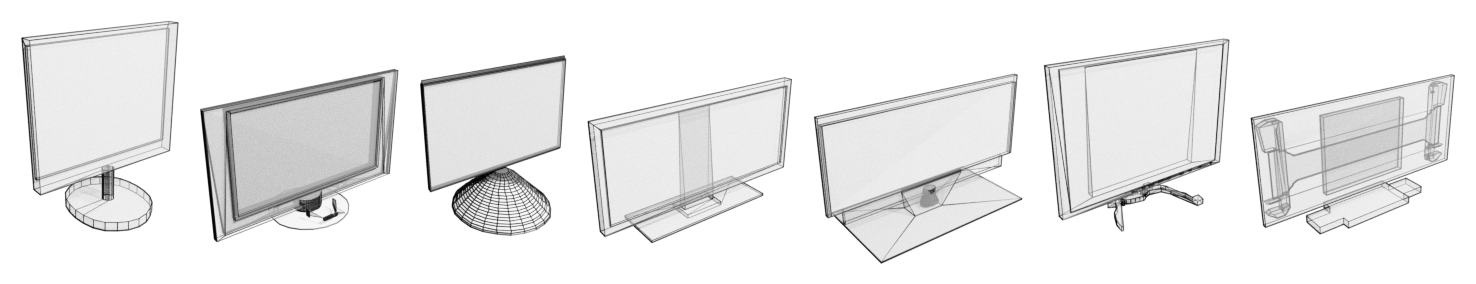}%
    }
    \ULsubfloat[Chair]{%
      \includegraphics[width=0.92\linewidth]{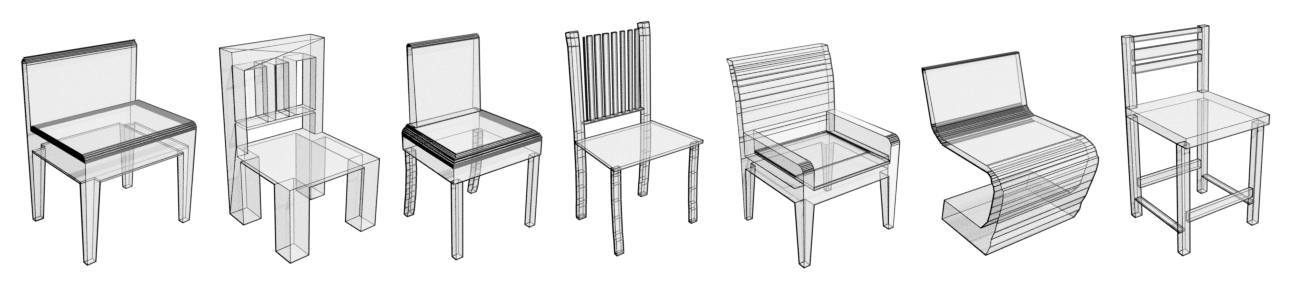}%
    }
    \ULsubfloat[Table]{%
       \includegraphics[width=0.92\linewidth]{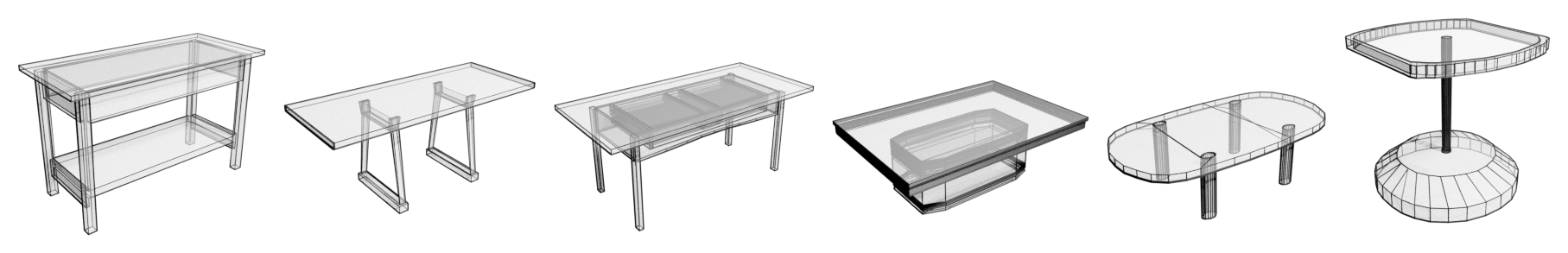}%
    }
    \caption{Class conditional samples generated by PolyGen using nucleus sampling and top-$p=0.9$.}
    \label{fig:class-conditional-samples}
\end{figure}

\boldpar{Vertices as discrete variables} We apply 8-bit uniform quantization to the mesh vertices. This reduces the size of meshes as nearby vertices that fall into the same bin are merged. We model the quantized vertex values using a Categorical distribution, and output at each step the logits of the distribution. This approach has been used to model discretized continuous signals in PixelCNN \cite{DBLP:conf/icml/OordKK16}, and WaveNet \cite{DBLP:conf/ssw/OordDZSVGKSK16}, and has the benefit of being able to express distributions without shape limitations. Mesh vertices have strong symmetries and complex dependencies, so the ability to express arbitrary distributions is important. We find 8-bit quantization to be a good trade-off between mesh fidelity, and mesh size. However, it should be noted that 14-bits or higher is typical for lossy mesh compression, and in future work it would be desirable to extend our methods to higher resolution meshes.
\begin{figure}[t]
    \centering
    \includegraphics[width=\linewidth]{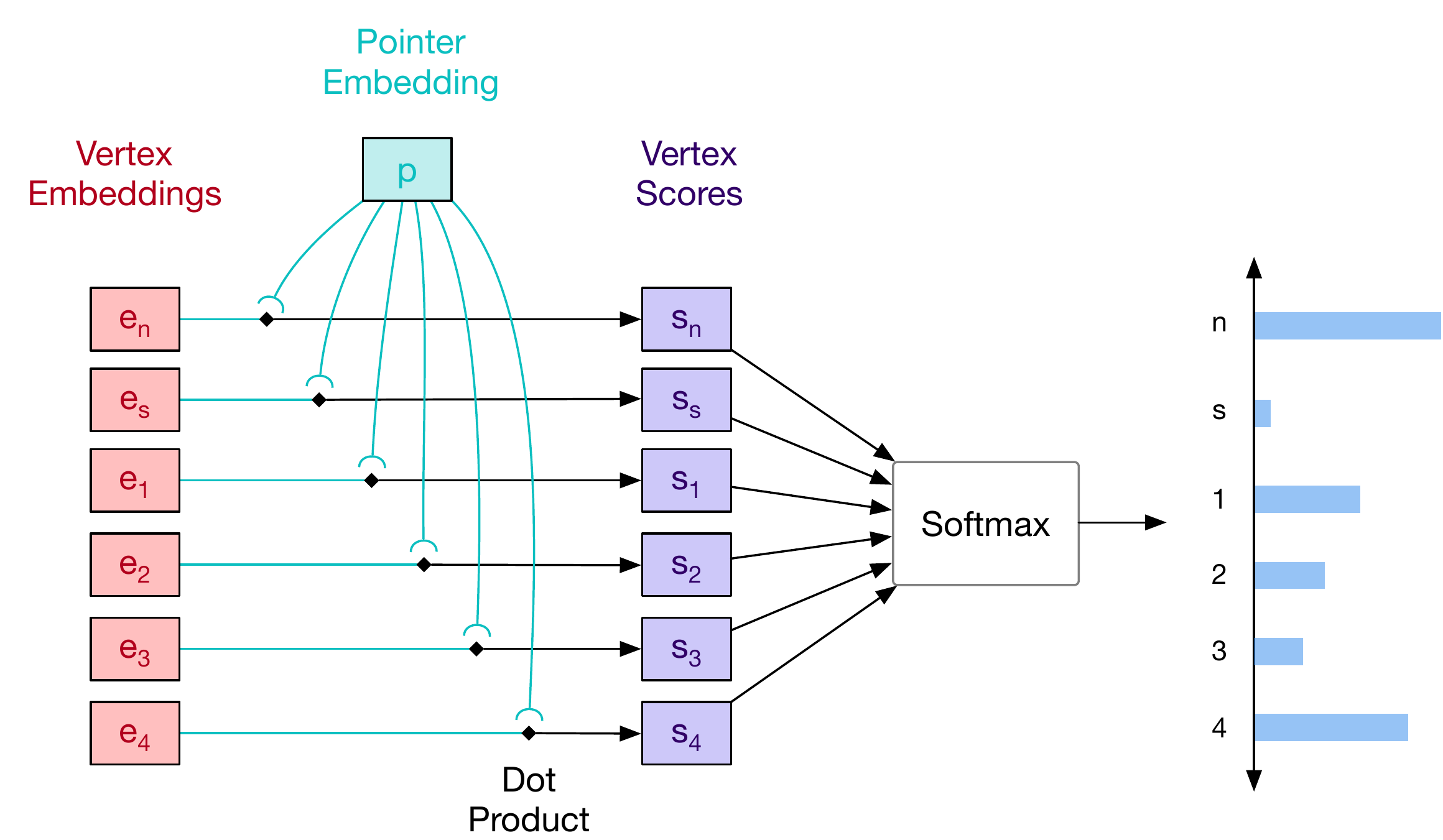}
    \caption{The mesh pointer network produces a distribution over variable length vertex sequences by comparing an output pointer embedding to
    vertex embeddings. In this example the number of vertices under consideration $N_V=4$ and therefore the distribution is over 6 elements.}
    \label{fig:pointer}
\end{figure}

\boldpar{Embeddings} \label{subsec:vertex-model-embeddings} We found the approach of using learned position and value embedding methods proposed in \cite{DBLP:journals/corr/abs-1904-10509} to work well. We  use three embeddings for each input token: A coordinate embedding, that indicates whether the input token is an $x$, $y$, or $z$ coordinate, a position embedding, that indicates which vertex in the sequence the token belongs to, and a value embedding, which expresses a token's quantized coordinate value. We use learned discrete embeddings in each case. 

\boldpar{Improving efficiency} \label{subsec:vertex-model-efficiency} One of the downsides of using Transformers for modelling sequential data is that they incur significant computational costs due to the quadratic nature of the attention operation. This presents issues when it comes to scaling our models to larger meshes. To address this, we explored several modifications of the model inspired by \cite{Salimans17}. All of them relieve the computational burden by chunking the sequence into triplets of vertex coordinates and processing each of them at once. The first variant uses a mixture of discretized logistics to model whole 3D vertices. The second replaces the mixture with a MADE-based decoder \citep{Germain15}. Finally, we present variants that use a Transformer decoder but rely on different vertex embedding schemes. These modifications are described in more detail in appendix E.

\subsection{Face Model}
\label{subsec:face-model}
The face model expresses a distribution over a sequence of mesh faces conditioned on the mesh vertices.  We order the faces by their lowest vertex index, then by their next lowest vertex and so on, where the vertices have been ordered from lowest to highest as described in Section \ref{subsec:vertex-model}.  Within a face we cyclically permute the face indices so that the lowest index is first. As with the vertex sequences, we concatenate the faces $(f_1^{(i)}, f_2^{(i)}, \dotsc, f_{N_i}^{(i)})_i$ to form a flattened sequence, with a final stopping token. We write $\mathcal{F}^{\textrm{seq}}$ for this flattened sequence, with elements $f_n, n=1, \dotsc, N_F$. 
\begin{align}
    p(\mathcal{F}^{\textrm{seq}} | \mathcal{V} ; \theta)  = \prod_{n=1}^{N_F} p(f_n | f_{<n}, \mathcal{V} ; \theta)
\end{align}
As with the vertex model, we output a distribution over the values of $f$ at each step, and train by maximizing the log-likelihood of $\theta$ over the training set. The distribution is a categorical defined over $\{1, \dotsc, N_V + 2\}$ where $N_V$ is the number of input vertices, and we include two additional values for the end-face $n$ and stopping $s$ tokens.  

\boldpar{Mesh pointer networks}
The target distribution $p(f_n | f_{<n}, \mathcal{V} ; \theta)$ is defined over the indices of an input set of vertices, which poses the challenge that the size of this set varies across examples. Pointer networks \cite{DBLP:conf/nips/VinyalsFJ15} propose an elegant solution to this issue; Firstly the input set is embedded using an encoder, and then at each step an autoregressive network outputs a pointer vector that is compared to the input embeddings via a dot-product. The resulting scores are then normalized using a softmax to form a valid distribution over the input set.

In our case we obtain contextual embeddings $\vec{e}_v$ of the input vertices using a Transformer encoder $E$. This has the advantage of bi-directional information aggregation compared to the LSTM used by the original pointer networks. We jointly embed new-face and stopping tokens with the vertices, to obtain a total of $N_V + 2$ input embeddings. A Transformer decoder $D$ operates on the sequence of faces and outputs pointers $\vec{p}_k$ at each step. The target distribution can be obtained as
\begin{align}
    \{\vec{e}_v\}_{v=1}^{N_V} &= E(\mathcal{V} ; \theta) \\
    \vec{p}_n &= D(f_{<n}, \mathcal{V} ; \theta)\\
    p(f_n = k \ | \ f_{<n}, \mathcal{V} ; \theta) &= \text{softmax}_k(\vec{p}_n^T\vec{e}_k)
\end{align}
See Figure~\ref{fig:pointer} for an illustration of the pointer mechanism and Figure~\ref{fig:mesh-model} in the appendix for an illustration of the whole face model. The decoder $D$ is a masked Transformer decoder that operates on sequences of embedded face tokens. It conditions on the input vertices in two ways, via dynamic face embeddings as explained in the next section, and optionally through cross-attention into the sequence of vertex embeddings. 

\boldpar{Embeddings} \label{subsec:face-model-embeddings} As with the vertex model we use learned position and value embeddings. We decompose a token's position into the index of the face it belongs to, as well as the location of a token within a face, using separate learned embeddings for both. For value embeddings we follow the approach of pointer networks and simply embed the vertex indices by indexing into the contextual vertex embeddings outputted by the vertex encoder. 

\subsection{Masking Invalid Predictions} 
\label{subsec:masking}
For both the vertex and face model only certain predictions are valid at each step. For instance, the $z$-coordinates must increase monotonically, and the stopping token can only be placed after an $x$ coordinate. Similarly mesh faces can not have duplicate indices, and every vertex-index must be referenced by at least one face. When evaluating the model we mask the predicted logits to ensure that the model can only make valid predictions. This has a non-negative effect on the model's log-likelihood scores, as it reassigns probability mass in the invalid region to values in the valid region (Table~\ref{table:log-likelihood}). Surprisingly, we found that masking during training to worsen performance to a small degree, so we always train without masking. For a complete description of the masks used, see appendix F.
\begin{figure}[t]
    \centering
    \begin{subfigure}[b]{\linewidth}
        \centering
        \includegraphics[width=\linewidth]{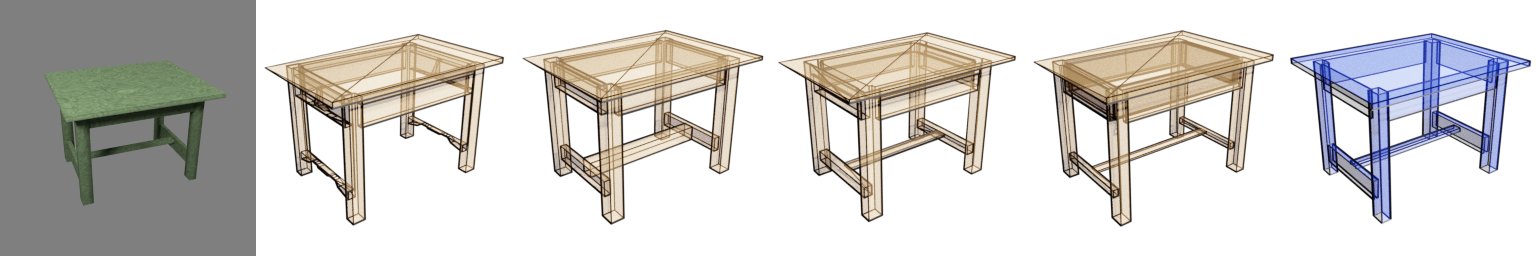}
    \end{subfigure}
    \begin{subfigure}[b]{\linewidth}
        \centering
        \includegraphics[width=\linewidth]{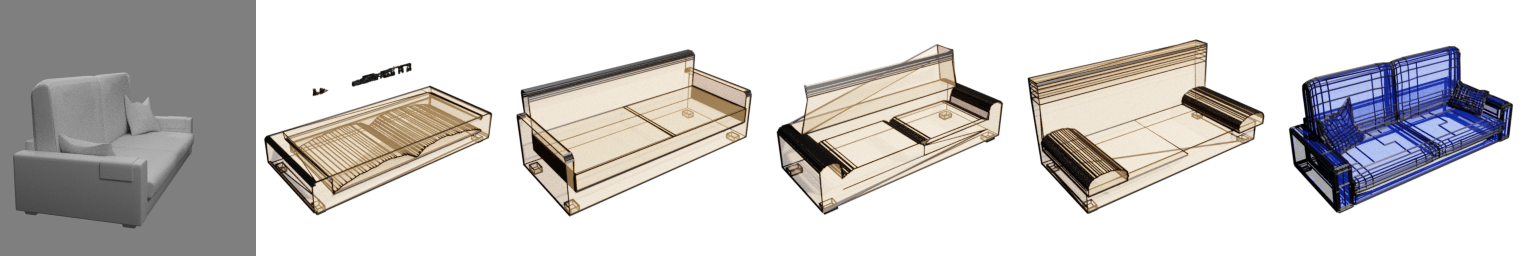}
    \end{subfigure}
    \begin{subfigure}[b]{\linewidth}
        \centering
        \includegraphics[width=\linewidth]{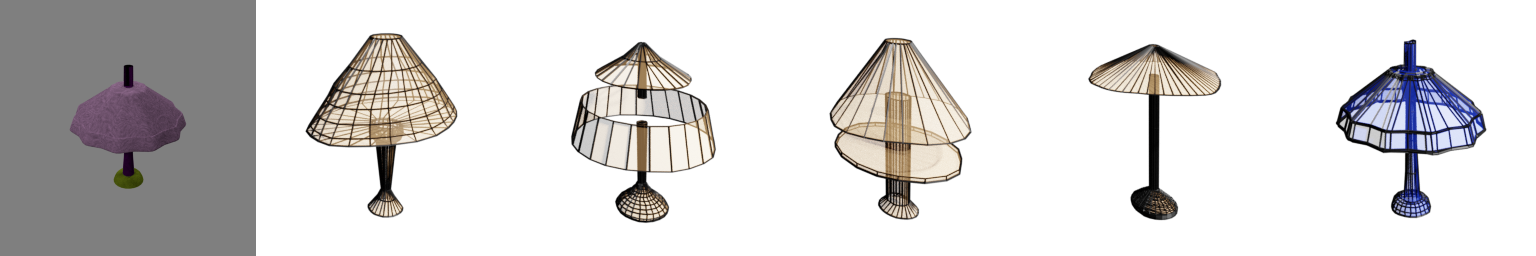}
    \end{subfigure}
    \begin{subfigure}[b]{\linewidth}
        \centering
        \includegraphics[width=\linewidth]{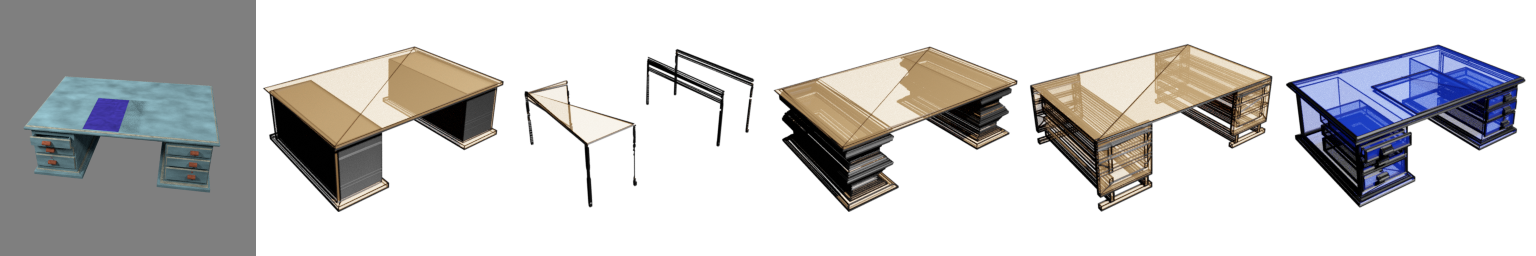}
    \end{subfigure}
    \begin{subfigure}[b]{\linewidth}
        \centering
        \includegraphics[width=\linewidth]{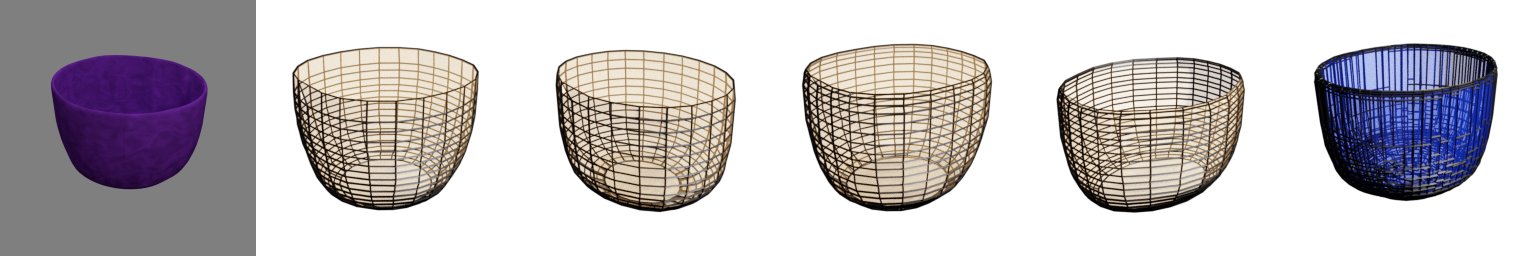}
    \end{subfigure}
    \caption{Image conditional samples (yellow) generated using nucleus sampling with top-$p$=0.9 and ground truth meshes (blue).}
    \label{fig:image-conditional}
\end{figure}
\begin{table}[t]
    \caption{Modelling performance of unconditional models trained on ShapeNet and baseline methods. Negative log-likelihood is reported in bits per vertex, averaged across test examples. Accuracy refers to the classification accuracy of next step predictions: discrete vertex coordinates for the vertex model, or vertex indices for face models. *Draco is evaluated on triangulated meshes rather than $n$-gon meshes.}
    \label{table:log-likelihood}
    \centering
    \begin{tabular}{@{}lcccc@{}}
        \toprule
                                                  & \multicolumn{2}{c}{Bits per vertex} & \multicolumn{2}{c}{Accuracy} \\ \cmidrule(lr){2-3} \cmidrule(lr){4-5}
    Model & Vertices & Faces & Vertices & Faces \\ 
    \midrule Uniform & 24.08 & 39.73 & 0.004 & 0.002 \\
    Valid predictions & 21.41 & 25.79 & 0.009 & 0.038 \\
    Draco* \cite{draco} & \multicolumn{2}{c}{Total: 27.68} & - & - \\
        \midrule PolyGen & 2.46 & 1.79 & 0.851 & 0.900 \\ 
    {\footnotesize - valid predictions} & 2.47 & 1.82 & 0.851 & 0.900 \\ 
    {\footnotesize - discr. embed. (V)} & 2.56 & - & 0.844 & - \\ 
    {\footnotesize - data augmentation} & 3.39 & 2.52 & 0.803 & 0.868 \\ 
                        
    {\footnotesize + cross attention (F)} & - & 1.87 & - & 0.899 \\
    \bottomrule
    \end{tabular}
    \end{table}
    
\begin{table}[t]
    \caption{Comparison of vertex model variants. The first two columns correspond to the test negative log-likelihood and predictive accuracy (see Table~\ref{table:log-likelihood}). The last column shows training speed in steps per second.}
    \label{table:alternative-vertex-models}
    \centering
    \begin{tabular}{@{}lccc@{}}
        \toprule
        Model & \shortstack{Bits \\ {\small per vertex}} & Accuracy & \shortstack{Steps \\ {\small per sec}} \\
        \midrule
        Mixture & 3.01 & - & 7.19 \\
        MADE decoder & 2.65 & 0.844 & 7.02 \\
        Tr. decoder & 2.50 & 0.851 & 4.07 \\
        {\footnotesize + Tr. embed.} & 2.48 & 0.851 & 4.60 \\
        \midrule
        Base model & 2.46 & 0.851 & 2.98 \\ 
        \bottomrule
    \end{tabular}
\end{table}
\subsection{Conditional Mesh Generation} 
\label{sec:conditional}
We can guide the generation of mesh vertices and faces by conditioning on a context. For instance, we can output vertices consistent with a given object class, or infer the mesh associated with an input image. It is straightforward to extend the vertex and face models to condition on a context $\vec{h}$. We incorporate context in two ways, depending on the domain of the input. For global features like class identity, we project learned class embeddings to a vector that is added to the intermediate Transformer representations following the self-attention layer in each block. For high dimensional inputs like images, or voxels, we jointly train a domain-appropriate encoder that outputs a sequence of context embeddings. The Transformer decoder then performs cross-attention into the embedding sequence, as in the original machine translation Transformer model.

For image inputs we use an encoder consisting of a series of downsampling residual blocks. We use pre-activation residual blocks \cite{DBLP:conf/eccv/HeZRS16}, and downsample three times using convolutions with stride 2, taking input images of size $[256,256,3]$ to feature maps of size $[16,16,E]$ where $E$ is the embedding dimensionality of the model. For voxel inputs we use a similar encoder but with 3D convolutions that takes inputs of shape $[28,28,28,1]$ to spatial embeddings of shape $[7,7,7,E]$. For both input types we add coordinate embeddings to the feature maps before flattening the spatial dimensions. For more architecture details see appendix C.

\section{Experiments} \label{experiments} 
Our primary evaluation metric is log-likelihood, which we find to correlate well with sample quality. We also report summary statistics for generated meshes, and compare our model to existing approaches using chamfer-distance in the image and voxel conditioned settings.

\subsection{Training Details} \label{subsec:training}
We train all our models on the ShapeNet Core V2 dataset \cite{shapenet2015}, which we subdivide into $92.5\%$ training, $2.5\%$ validation and $5\%$ testing splits. The training set is augmented as described in Section \ref{subsec:augmentation}. In order to reduce the memory requirements of long sequences we filter out meshes with more than 800 vertices, or more than 2800 face indices after pre-processing. We train the vertex and face models for $1e6$ and $5e5$ weight updates respectively, using four V100 GPUs per training run for a total batch size of 16. We use the Adam optimizer with a gradient clipping norm of 1.0, and perform cosine annealing from a maximum learning rate of $3e-4$, with a linear warm up period of 5000 steps. We use a dropout rate of 0.2 for all models.

\subsection{Data Augmentation and Rendering} \label{subsec:augmentation}
In general we observed significant overfitting due to the relatively small size of the ShapeNet dataset, which is exacerbated by the need to filter out large meshes. In order to reduce this effect, we augmented the input meshes by scaling the vertices independently on each axis, using a random piecewise-linear warp for each axis, and by varying the decimation angle used to create $n$-gon meshes. For each input mesh we create 50 augmented versions which are then quantized (Section \ref{subsec:vertex-model}) for use during training. We found that augmentation was necessary to obtain good performance (Table~\ref{table:log-likelihood}). For full details of the augmentations and parameter settings see appendix A.

\boldpar{Rendering} In order to train image-conditional models we create renders of the processed ShapeNet meshes using Blender \cite{Blender}. For each augmented mesh, and each validation and test-set mesh, we create renders at $256 \times 256$ resolution, using randomly chosen lighting, camera and mesh material settings. For more details see appendix B.

\subsection{Unconditional Modelling Performance}
We compare unconditional models trained under varying conditions. As evaluation metrics we report the negative log-likelihood obtained by the models, reported in bits per vertex, as well as the accuracy of next step predictions. For vertex models this is the accuracy of next vertex coordinate predictions, and for face models this is the accuracy of the next vertex index predictions. 
In particular we compare the effect of masking invalid predictions (Section \ref{subsec:masking}), of using discrete rather than continuous coordinate embeddings in the vertex model (Section \ref{subsec:vertex-model-embeddings}), of using data augmentation (Section \ref{subsec:augmentation}), and finally of using cross-attention in the face model. Unless otherwise specified we use embeddings of size 256, fully connected layers of size 1024, and 18 and 12 Transformer blocks for the vertex and face models respectively. As there are no existing methods that directly model mesh vertices and faces, we report the scores obtained by models that allocate uniform probability to the whole data domain, as well as models that are uniform over the region of valid predictions. We additionally report the compression rate obtained by Draco \cite{draco}, a mesh compression library. For details of the Draco compression settings see appendix G.

Table~\ref{table:log-likelihood} shows the results obtained by the various models. We find that our models achieve significantly better modelling performance than the uniform and Draco baselines, which illustrates the gains achievable by a learned predictive model. We find that restricting the models predictions to the range of valid values results in a minor improvement in modelling performance, which indicates that the model is effective at assigning low probability to the invalid regions. Using discrete rather than continuous embeddings for vertex coordinates provides a significant improvement, improving bits-per-vertex from 2.56 to 2.46. Surprisingly, using cross-attention in the face model harms performance, which we attribute to overfitting. Data augmentation has a strong effect on performance, with models trained without augmentation losing 1.64 bits per vertex on average. Overall, our best model achieves a log-likelihood score of 4.26 bits per vertex, and $85\%$ and $90\%$ predictive accuracy for the vertex and face models respectively. Figure \ref{fig:unconditional} in the appendix shows random unconditional samples from the best performing model.

Table~\ref{table:alternative-vertex-models} presents a comparison of different variants of the vertex model as discussed in Section \ref{subsec:vertex-model-efficiency}. The results suggest that the proposed variants can achieve a $ 1.5\times $ reduction in training time with a minimal sacrifice in performance. Note that these models used different hyperparameter settings as detailed in Appendix E.
\begin{figure}[t]
    \centering
    \includegraphics[width=\linewidth]{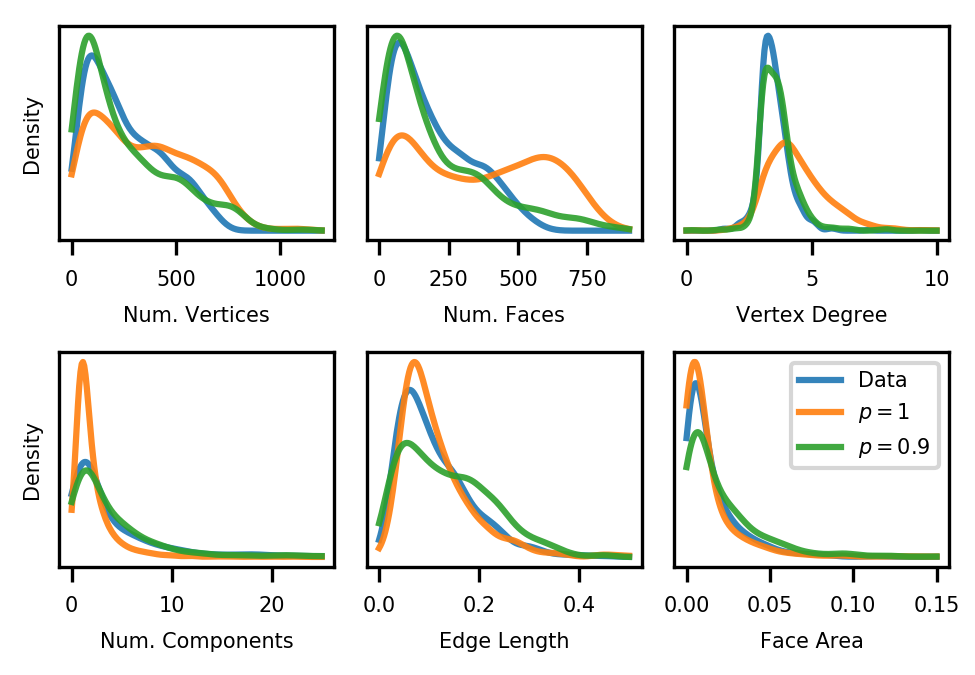}
    \caption{Distribution of mesh statistics for unconditional samples from our model and the ShapeNet test set. We compare samples generated using with nucleus sampling and top-$p=0.9$, to true model samples ($p=1$).}
    \label{fig:sample-statistics}
\end{figure}
\subsection{Statistics of Unconditional Model Samples}
We compare the distribution of certain mesh summaries for samples from our model against the ShapeNet test set.  If our model has closely matched the true data distribution then we expect these summaries to have similar distributions. We draw 1055 samples from our best unconditional model, and discard samples that don't produce a stopping token within 1200 vertices, or 800 faces. We use nucleus sampling \cite{DBLP:journals/corr/abs-1904-09751} which we found to be effective at maintaining sample diversity while reducing the presence of degraded samples. Nucleus sampling helps to reduce sampling degradation by sampling from the smallest subset of tokens that account for top-$p$ of probability mass.

\begin{figure}[t]
    \centering
    \begin{subfigure}[b]{0.495\linewidth}
        \centering
        \includegraphics[width=\linewidth]{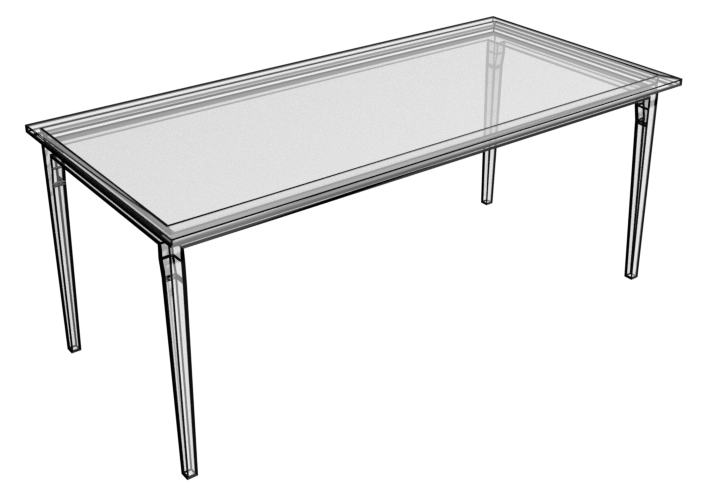}
        \caption{PolyGen}
    \end{subfigure}
    \begin{subfigure}[b]{0.495\linewidth}
        \centering
        \includegraphics[width=\linewidth]{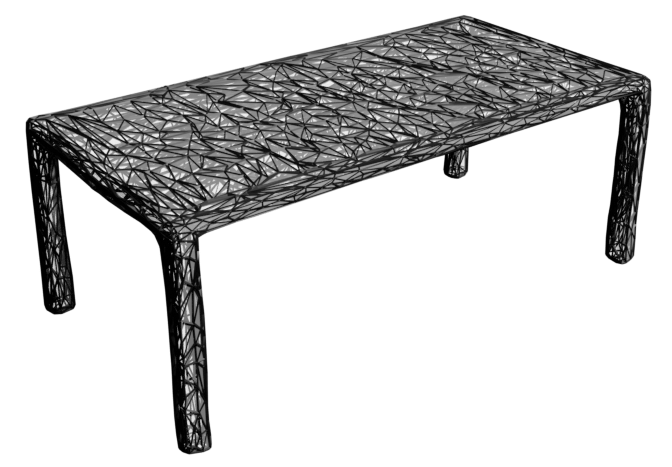}
        \caption{Occupancy Networks}
    \end{subfigure}

    \caption{Comparison between a 3D mesh generated by PolyGen and a mesh obtained by postprocessing an implicit surface representation (Occupancy Networks \citealp{DBLP:conf/cvpr/MeschederONNG19}). Our model produces a more efficient representation of the 3D shape that resembles a human-constructed mesh.}
    \label{fig:occ-comparison}
\end{figure}

Figure~\ref{fig:sample-statistics} shows the distribution of a number of mesh summaries, for samples from PolyGen as well as the true data distribution. In particular we show: the number of vertices, number of faces, node degree, average face area and average edge length for sampled and true meshes. Although these are coarse descriptions of a 3D mesh, we find our model's samples to have a similar distribution for each mesh statistic. We observe that nucleus sampling with $\text{top-}p = 0.9$ helps to align the model distributions with the true data for a number of statistics. Figure \ref{fig:occ-comparison} shows an example 3D mesh generated by our model compared to a mesh obtained through post-processing an occupancy function \cite{DBLP:conf/cvpr/MeschederONNG19}. We note that the statistics of our mesh resemble human-created meshes to a greater extent.
\begin{table}[t]
    \caption{Modelling performance for conditional models. See Table \ref{table:log-likelihood} for details of bits per vertex, and accuracy scores.}
    \label{table:conditional-log-likelihood}
    \centering
    \begin{tabular}{@{}lccccc@{}}
        \toprule
        & \multicolumn{3}{c}{Bits per vertex} & \multicolumn{2}{c}{Accuracy} \\ 
    \cmidrule(lr){2-4} \cmidrule(lr){5-6} Context & Vertices & Faces  & Total & Vertices & Faces\\ 
    \midrule None & 2.46 & 1.79 & 4.26 & 0.851 & 0.900  \\
    Class  & 2.43 & 1.81 & 4.24 & 0.853 & 0.899 \\
    Image  & 2.30 & 1.81 & 4.11 & 0.857 & 0.900 \\
    {\small  + pooling}  & 2.35 & 1.78 & 4.13 & 0.856 & 0.900 \\
    Voxels  & 2.19 & 1.82 & 4.01 & 0.859 & 0.900 \\ 
    {\small  + pooling}  & 2.28 & 1.79 & 4.07 & 0.856 & 0.900 \\ \bottomrule
    \end{tabular}
\end{table}
\subsection{Conditional Modelling Performance}
\begin{figure}[t]
    \centering
    \begin{subfigure}[b]{\linewidth}
        \centering
        \includegraphics[width=\linewidth]{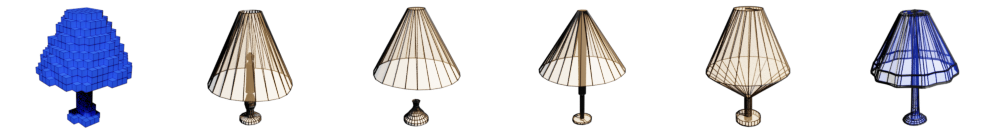}
    \end{subfigure}
    \vspace{5pt}
    \begin{subfigure}[b]{\linewidth}
        \centering
        \includegraphics[width=\linewidth]{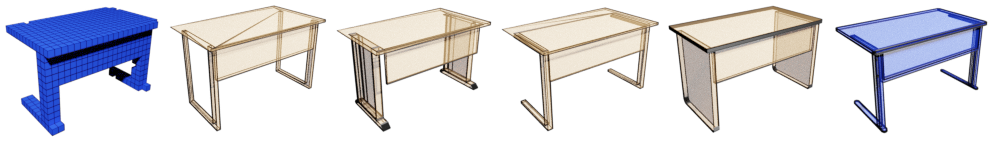}
    \end{subfigure}
    \vspace{5pt}
    \begin{subfigure}[b]{\linewidth}
        \centering
        \includegraphics[width=\linewidth]{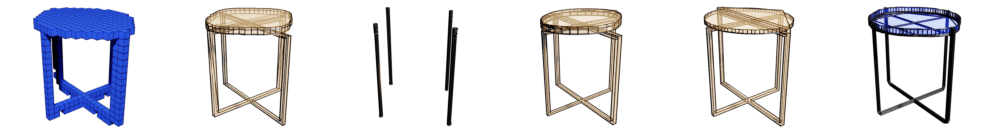}
    \end{subfigure}
    \vspace{5pt}
    \begin{subfigure}[b]{\linewidth}
        \centering
        \includegraphics[width=\linewidth]{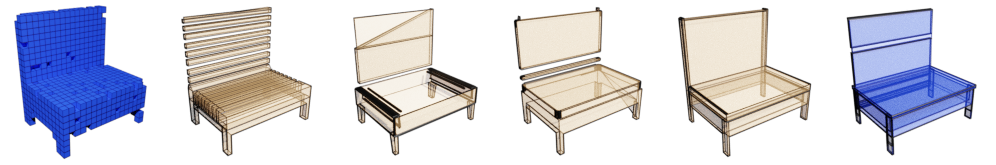}
    \end{subfigure}
    \vspace{5pt}
    \begin{subfigure}[b]{\linewidth}
        \centering
        \includegraphics[width=\linewidth]{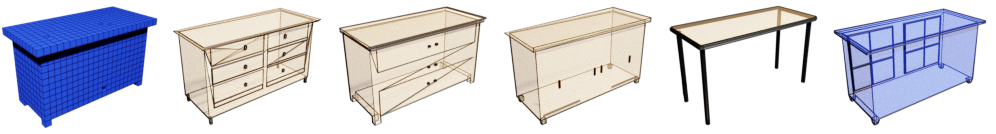}
    \end{subfigure}
    \caption{Voxel conditional (blue, left) samples generated using nucleus sampling with top-$p$=0.9 (yellow) and ground truth meshes (blue, right).}
    \label{fig:voxel-conditional}
\end{figure}

We train vertex and face models with three kinds of conditioning: class labels, images, and voxels. We use the same settings as the best unconditional model: discrete vertex embeddings with no cross attention in the face model. As with the unconditional models we use 18 layers for the vertex model and 12 layers for the face model. Figures \ref{fig:teaser} and \ref{fig:class-conditional-samples} show class-conditional samples. Figures \ref{fig:image-conditional} and \ref{fig:voxel-conditional} show samples from image and voxel conditional models respectively. Note that while we train on the ShapeNet dataset, we show ground truth meshes and inputs for a selection of representative meshes collected from the \citet{turbosquid} online object repository.
\begin{figure}[t]
    \centering
    \includegraphics[width=\linewidth]{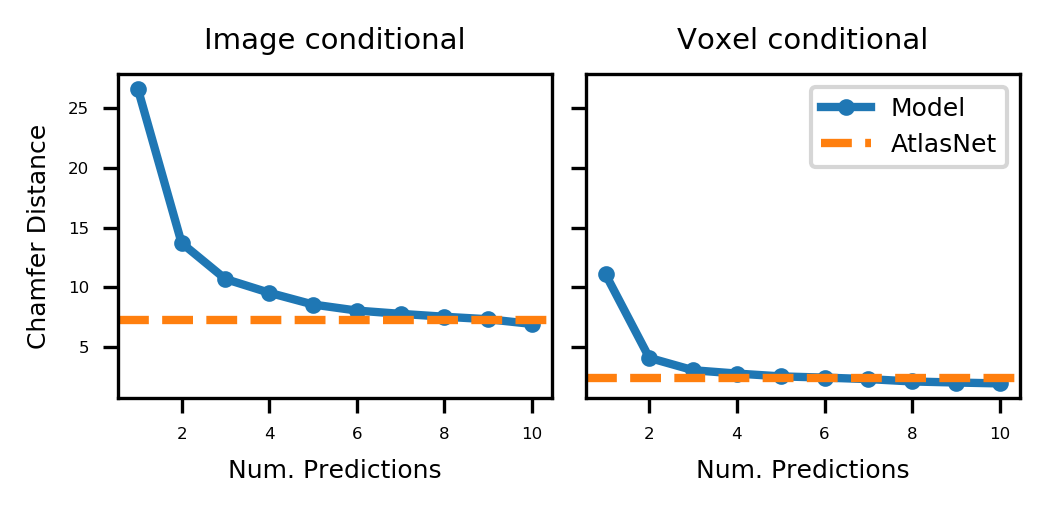}
    \caption{Symmetric chamfer distance between predicted and target pointclouds by number of predictions. Data refers to pointclouds obtained by uniformly re-sampling the target mesh. }
    \label{fig:chamfer}
\end{figure}
Table~\ref{table:conditional-log-likelihood} shows the impact of conditioning on predictive performance in terms of bits-per-vertex and accuracy. We find that for vertex models, voxel conditioning provides the greatest improvement, followed by images, and then by class labels. This confirms our expectations, as voxels characterize the coarse shape unambiguously, whereas images can be ambiguous depending on the object pose and lighting. However the additional context does not lead to improvements for the face model, with all conditional face models performing slightly worse than the best unconditional model. This is likely because mesh faces are to a large extent determined by the input vertices, and the conditioning context provides relatively little additional information. In terms of predictive accuracy, we see similar effects, with accuracy improving with richer contexts for vertex models, but not for face models. We note that the accuracy ceiling is less than $100\%$, due to the inherent entropy of the vertex and face distributions, and so we expect diminishing gains as models approach this ceiling. 

For image and voxel conditional models, we also compare to architectures that apply global average pooling to the outputs of the input encoders. We observe that pooling in this way negatively effects the vertex models' performance, but has a small positive effect on the face models' performance. 

\subsection{Mesh Reconstruction}
We additionally evaluate the image and voxel conditioned models on mesh reconstruction, where we use symmetric chamfer distance as the reconstruction metric. The symmetric chamfer distance is a distance metric between two point sets $\mathcal{P}$ and $\mathcal{Q}$. It is defined as:
\begin{align}
    \mathcal{L}(\mathcal{P}, \mathcal{Q}) = \sum_{\vec{p} \in \mathcal{P}} \underset{\vec{q} \in \mathcal{Q}}{\text{min}} (\vec{p} - \vec{q})^2 + \sum_{\vec{q} \in \mathcal{Q}} \underset{\vec{p} \in \mathcal{P}}{\text{min}} (\vec{p} - \vec{q})^2
\end{align}
For each example in the test set we draw samples from the conditional model. We sample 2500 points uniformly on the sampled and target mesh and compute the corresponding chamfer distance. We compare our model to AtlasNet \cite{DBLP:conf/cvpr/GroueixFKRA18}, a conditional model that defines a mesh surface using a number of patches that have been Transformer using a deep network. AtlasNet outputs pointclouds and is trained to minimize the chamfer distance to a target pointcloud conditioned on image or pointcloud inputs. Compared to alternative methods, AtlasNet achieves good mesh reconstruction performance, and we therefore view it as a strong baseline. We train AtlasNets models in the image and voxel conditioned settings, that are adapated to use equivalent image and voxel encoders as we use for our model. For more details see appendix D.

Figure~\ref{fig:chamfer} shows the mesh reconstruction results. We find that when making a single prediction, our model performs worse than AtlasNet. This is not unexpected, as AtlasNet optimizes the evaluation metric directly, whereas our model does not. When allowed to make 10 predictions, our model achieves slightly better performance than AtlasNet. Overall we find that while our model does not always produce good mesh reconstructions, it typically produces a very good reconstruction within 10 samples, which may be sufficient for many practical applications.

\section{Related Work}
\label{related-work}
Generative models of 3D objects  exists in a variety of forms, including ordered \cite{DBLP:journals/cgf/NashW17} and unordered \cite{DBLP:conf/iclr/LiZZPS19, DBLP:journals/corr/abs-1906-12320} pointclouds, voxels \cite{DBLP:conf/eccv/ChoyXGCS16, DBLP:conf/nips/0001ZXFT16, DBLP:conf/iccv/TatarchenkoDB17, DBLP:conf/nips/RezendeEMBJH16}. More recently there has been significant progress using functional representations, such as signed distance functions \cite{DBLP:conf/cvpr/ParkFSNL19}, and other implicit functions \cite{DBLP:conf/cvpr/MeschederONNG19}. There are relatively fewer examples of methods that explicitly generate a 3D mesh. Such works primarily use parameterized deformable meshes \cite{DBLP:conf/cvpr/GroueixFKRA18}, or form meshes through a collection of mesh patches.  Our methods are distinguished in that we directly model the mesh data created by people, rather than alternative representations or parameterizations. In addition, our model is probabilistic, which means we can produce diverse output, and respond to ambiguous inputs in a principled way. 

PolyGen's vertex model is similar to PointGrow \cite{sun2018pointgrow}, which uses an autoregressive decomposition to model 3D point clouds, outputting discrete coordinate distributions using a self-attention based architecture. PointGrow operates on fixed-length point-clouds rather than variable vertex sequences, and uses a bespoke self-attention architecture, that is relatively shallow in comparison to modern autoregressive models in other domains. By contrast, we use state-of-the-art deep architectures, and model vertices and faces, enabling us to generate high quality 3D meshes.

This work borrows from architectures developed for sequence modelling in natural language processing. This includes the sequence to sequence training paradigm \cite{DBLP:conf/nips/SutskeverVL14}, the Transformer architecture \cite{DBLP:conf/nips/VaswaniSPUJGKP17, DBLP:journals/corr/abs-1904-10509, DBLP:journals/corr/abs-1910-06764}, and pointer networks \cite{DBLP:conf/nips/VinyalsFJ15}. In addition our work is inspired by sequential models of raw data, like WaveNet \cite{DBLP:conf/ssw/OordDZSVGKSK16} PixelRNN and its variants \cite{DBLP:conf/nips/OordKEKVG16, DBLP:conf/iclr/MenickK19}, and Music Transformers \cite{DBLP:conf/iclr/HuangVUSHSDHDE19}. 

Our work is also related to Polygon-RNN \cite{DBLP:conf/cvpr/CastrejonKUF17, DBLP:conf/cvpr/AcunaLKF18}, a method for efficient segmentation in computer vision using polygons. Polygon-RNN take an input image and autoregressively outputs a sequence of $xy$ coordinates that implicitly define a segmented region. PolyGen, by contrast operates in 3D space, and explicitly defines the connectivity of several polygons. 

Finally our work is related to generative models of graph structured data such as GraphRNN \cite{DBLP:conf/icml/YouYRHL18} and GRAN \cite{DBLP:journals/corr/abs-1910-00760}, in that meshes can be thought of as attributed graphs. These works focus on modelling graph connectivity rather than graph attributes, whereas we model both the node attributes (vertex positions), as well as the incorporating these attributes in our model of the connectivity. 

\section{Conclusion}
\label{conclusion}
In this work we present PolyGen, a deep generative model of 3D meshes. We pose the problem of mesh generative as autoregressive sequence modelling, and combine the benefits of Transformers and pointer networks in order to flexibly model variable length mesh sequences. PolyGen is capable of generating coherent and diverse mesh samples, and we believe that it will unlock a range of applications in computer vision, robotics, and  3D content creation. 

\section*{Acknowledgements}


The authors thank Dan Rosenbaum, Sander Dieleman, Yujia Li and Craig Donner for useful discussions·

\bibliography{bibliography}
\bibliographystyle{icml2020}

\clearpage
\section*{Appendix}

\begin{figure}[h!]
\centering
\includegraphics[width=\linewidth]{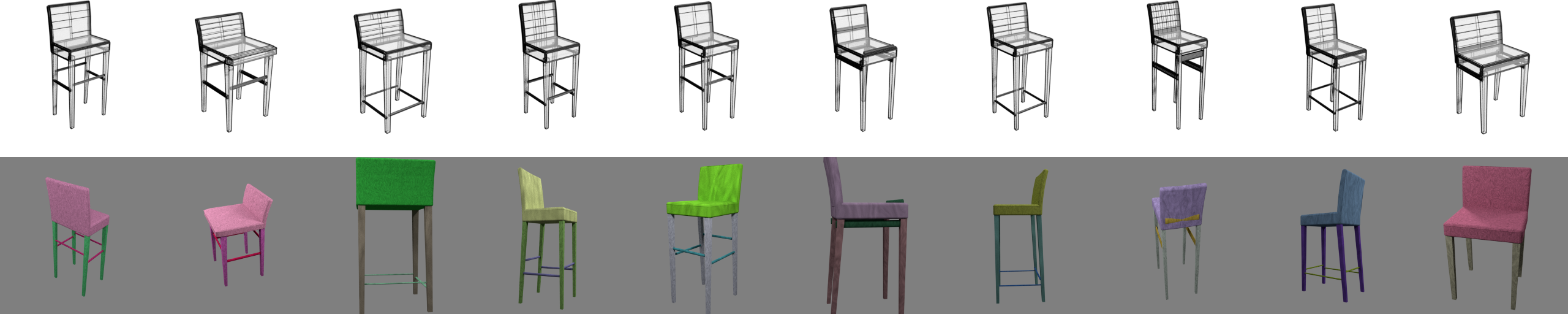}
\caption{Examples of data augmentation and randomized rendering conditions. For each input mesh we create 50 augmentations, and render each while varying lighting, camera and material properties.}
\label{fig:augs}
\end{figure}

\section*{A. Data Augmentation}
For each input mesh from the ShapeNet dataset we create 50 augmented versions which are used during training (Figure~\ref{fig:augs}). We start by normalizing the meshes such that the length of the long diagonal of the mesh bounding box is equal to 1. We then apply the following augmentations, performing the same bounding box normalization after each. All augmentations and mesh rendering are performed prior to vertex quantization.

\boldpar{Axis scaling} We scale each axis independently, uniformly sampling scaling factors $s_x$, $s_y$ and $s_z$ in the interval $[0.75, 1.25]$.

\boldpar{Piecewise linear warping} We define a continuous, piecewise linear warping function by dividing the interval $[0, 1]$ into 5 even sub-intervals, sampling gradients $g_1, \dotsc, g_5$ for each sub-interval from a log-normal distribution with variance 0.5, and composing the segments. For $x$ and $y$ coordinates, we ensure the warping function is symmetric about zero, by reflecting a warping function with three sub-intervals on $[0, 0.5]$ about 0.5. This preserves symmetries in the data which are often present for these axes.

\boldpar{Planar mesh decimation} We use Blender's planar decimation modifier (\url{https://docs.blender.org/manual/en/latest/modeling/modifiers/generate/decimate.html}) to create $n$-gon meshes. This merges adjacent faces where the angle between surfaces is greater than a certain tolerance. Different tolerances result in meshes of different sizes with differing connectivity due to varying levels of decimation. We use this property for data augmentation and sample the tolerance degrees uniformly from the interval $[1, 20]$ . 

\section*{B. Rendering}
We use Blender to create rendered images of the 3D meshes in order to train image-conditional models (Figure~\ref{fig:augs}). We use Blender's Cycles (\url{https://docs.blender.org/manual/en/latest/render/cycles/index.html}) path-tracing renderer, and randomize the lighting, camera, and mesh materials. In all scenes we place the input meshes at the origin, scaled so that bounding boxes are 1m on the long diagonal.

\boldpar{Lighting} We use an 20W area light located 1.5m above the origin, with rectangle size 2.5m, and sample a number of 15W point lights uniformly from the range $[0, 1, \dotsc, 10]$. We choose the location of each point light independently, sampling the $x$ and $y$ coordinates uniformly in the intervals $[-2, -0.75] \cup [0.75, 2]$, and sampling the $z$ coordinate uniformly in the interval $[0.75, 2]$.

\boldpar{Camera} We position the camera at a distance $d$ from the center of the mesh, where $d$ is sampled uniformly from $[1.25, 1.5]$, at an elevation sampled between $[0, 1]$, and sample a rotation uniformly between $[0, 360]$. We sample a focal length for the camera in $[35, 36, \dotsc, 50]$. We also sample a filter size (\url{https://docs.blender.org/manual/en/latest/render/cycles/render_settings/film.html}) in $[1.5, 2]$, which adds a small degree of blur.

\boldpar{Object materials}  We found the ShapeNet materials and textures to be applied inconsistently across different examples when using Blender, and in many cases no textures loaded at all. Rather than use the inconsistent textures, we randomly generated materials for the 3D meshes, in order to produce a degree of visual variability. For each texture group in the mesh we sampled a new material. Materials were constructed by linking Blender nodes (\url{https://docs.blender.org/manual/en/latest/render/shader_nodes/introduction.html#textures}). In particular we use a noise shader with detail = 16, scale $= \sqrt{100 * u}, u \sim \mathcal{U}(0, 1)$, and scale draw from the interval $[0, 20]$. The noise shader is used as input to a color ramp node which interpolates between the input color, and white. The color ramp node then sets the color of a diffuse BSDF material \url{https://docs.blender.org/manual/en/latest/render/shader_nodes/shader/diffuse.html}, which is applied to faces within a texture group.

\begin{figure*}[h!]
    \centering
    \includegraphics[width=0.75\linewidth]{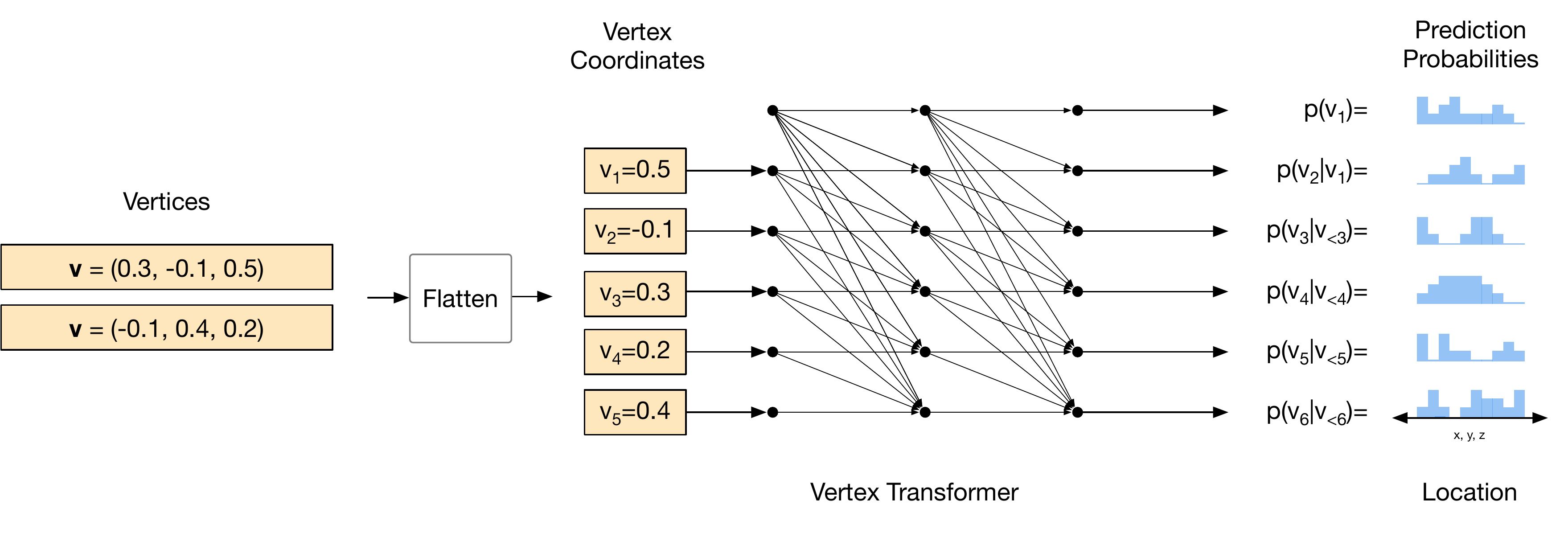}
    \caption{The vertex model is a masked Transformer decoder that takes as input a flattened sequence of vertex coordinates. The Transformer outputs discrete distributions over the individual coordinate locations, as well as the stopping token $s$. See Section~\ref{subsec:vertex-model} for a detailed description of the vertex model.}
    \label{fig:vertex-model}
\end{figure*}
\begin{figure*}[h!]
    \centering
    \includegraphics[width=0.8\linewidth]{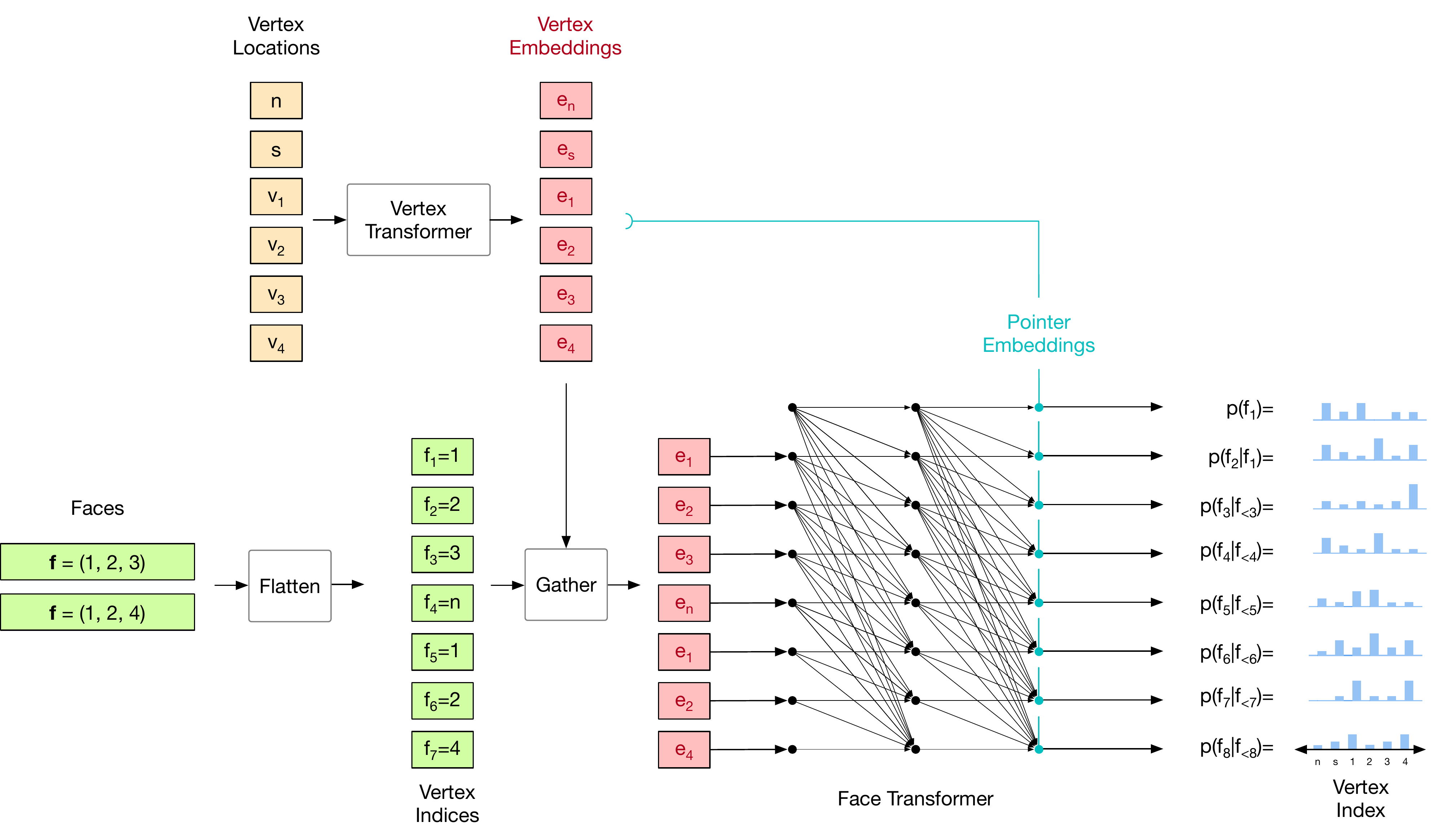}
    \caption{The face model model operates on an input set of vertices, as well as the flattened vertex indices that describe the faces. The vertices as well as the new face token $n$ and stopping token $s$ are first embedded using a Transformer encoder. A gather operation is then used to identify the embeddings associated with each vertex index. The index embeddings are processed with a masked Transformer decoder to output distributions over vertex indices at each step, as well as over the next-face token  and the stopping token. The final layer of the Transformer outputs pointer embeddings which are compared to the vertex embeddings using a dot-product to produce the desired distributions. See Section~\ref{subsec:face-model} for a detailed description of the face model and Figure~\ref{fig:pointer} in particular for a detailed depiction of the pointer network mechanism.}
    \label{fig:mesh-model}
\end{figure*}

\section*{C. Transformer blocks}
We use the improved Transformer variant with layer normalization moved inside the residual path, as in \cite{DBLP:journals/corr/abs-1904-10509, DBLP:journals/corr/abs-1910-06764}. In particular we compose the Transformer blocks as follows:
\begin{align}
    \vec{R}_{\text{MMH}}^{(l)} &= \text{MaskedMultiHead}(\text{LN}(\vec{H}_{\text{FC}}^{(l - 1)})) \\
    \vec{H}_{\text{MMH}}^{(l)} &= \vec{H}_{\text{FC}}^{(l - 1)} + \vec{R}_{\text{MMH}}^{(l)} \\
    \vec{R}_{\text{FC}}^{(l)} &= \text{Linear}(\text{ReLU}(\text{Linear}(\text{LN}(\vec{H}_{\text{MMH}}^{(l)})))) \\
    \vec{H}_{\text{FC}}^{(l)} &= \vec{H}_{\text{MMH}}^{(l)} + \vec{R}_{\text{FC}}^{(l)}
\end{align}
Where $\vec{R}^{(l)}$ and $\vec{H}^{(l)}$ are residuals and intermediate representations in the $l$'th block, and the subscripts FC and MMH denote the outputs of fully connected and masked multi-head self-attention layers respectively. We apply dropout immediately following the ReLU activation as this performed well in initial experiments.

\boldpar{Conditional models} 
As described in Section \ref{sec:conditional} For global features like class identity, we project learned class embeddings to a vector that is added to the intermediate Transformer representations $\vec{H}_{\text{MMH}}$ following the self-attention layer in each block:
\begin{align}
    \vec{r}_{\text{global}}^{(l)} &= \text{Linear}\left(\vec{h}_{\text{global}}\right) \\
    \vec{H}_{\text{global}}^{(l)} &= \vec{H}_{\text{MMH}}^{(l)} + \text{Broadcast}\left(\vec{r}_{\text{global}}^{(l)}\right)
\end{align}
For high dimensional inputs like images, or voxels, we jointly train a domain-appropriate encoder that outputs a sequence of context embeddings. The Transformer decoder performs cross-attention into the embedding sequence after the self-attention layer, as in the original machine translation Transformer model:
\begin{align}
    \vec{R}_{\text{seq}}^{(l)} &= \text{CrossMultiHead}\left(\vec{H}_{\text{MMH}}^{(l)}, \vec{H}_{\text{seq}}\right) \\
    \vec{H}_{\text{seq}}^{(l)} &= \vec{H}_{\text{MMH}}^{(l)} + \vec{R}_{\text{seq}}^{(l)}
\end{align}
The image and voxel encoders are both pre-activation resnets, with 2D and 3D convolutions respectively. The full architectures are described in Table \ref{tab:arch}. 

\section*{D. AtlasNet}
We use the same image and voxel-encoders (Table \ref{tab:arch}) as for the conditional PolyGen models. For consistency with the original method, we project the final feature maps to 1024 dimensions, before applying global average pooling to obtain a vector shape representation. As in the original method, the decoder is an MLP with 4 fully-connected layers of size 1024, 512, 256, 128 with ReLU non-linearities on the first three layers and tanh on the final output layer. The decoder takes the shape representation, as well as 2D points as input, and outputs a 3D vector. We use 25 patches, and train with the same optimization settings as PolyGen (Section \ref{experiments}) but for $5e5$ steps. 

\boldpar{Chamfer distance} To evaluate the chamfer distance for AtlasNet models, we first generate a mesh by passing 2D triangulated meshes through each of the AtlasNet patch models as described in \cite{DBLP:conf/cvpr/GroueixFKRA18}. We then sample points on the resulting 3D mesh.

\section*{E. Alternative Vertex Models}
In this section, we provide more details for the more efficient vertex model variants mentioned in Section \ref{subsec:vertex-model-efficiency}.

In the first variant, instead of processing $ x $, $ y $ and $ z $ coordinates in sequence we concatenate their embeddings together and pass them through a linear projection. This forms the input sequence for a $ 22 $-layer Transformer which we call \textit{the torso}. Following \cite{Salimans17} we output the parameters of a mixture of $ 40 $ discretized logistics describing the joint distribution of a full 3D vertex.  The main benefit of this model is that the self-attention is now performed for sequences which are $ 3 $ times shorter. This manifests in a much improved training time (see \ref{table:alternative-vertex-models}). Unfortunately, the speed-up comes at a price of significantly reduced performance. This may be because the underlying continuous components are not well suited to the peaky and multi-modal vertex distributions. 

In the second variant we lift the parametric distribution assumption and use a MADE-style masked MLP \citep{Germain15} with $ 2 $ residual blocks to decode each output of a $ 18 $-layer torso $ h_n $  into a sequence of three conditional discrete distributions:
\begin{align}
    p(v_n | h_n) &= p(z_n | h_n) p(y_n | z_n, h_n) p(x_n | z_n, y_n, h_n)
\end{align}
As expected, this change improves the test data likelihood while simultaneously increasing the computation cost. We notice that unlike the base model the MADE decoder has direct access only to the coordinate components within a single vertex and must rely on the output of the torso to learn about the components of previously generated vertices.

We let the decoder attend to all the generated coordinates directly in the third alternative version of our model. We replace the MADE decoder with a $ 6 $-layer Transformer which is conditioned on $ \{ h_n \}_n $ (this time produced by a $ 14 $-layer torso) and operates on a flattened sequence of vertex components (similarly to the base model). The conditioning is done by adding $ h_n $ to the embeddings of $ z_n $, $ y_n $ and $ x_n $. While slower than the MADE version, the resulting network is significantly closer in performance to the base model.

Finally, we make the model even more powerful using a $ 2 $-layer Transformer instead of simple concatenation to embed each triplet of vertex coordinates. Specifically, we sum-pool the outputs of that Transformer within every vertex. In this variant, we reduce the depth of the torso to $ 10 $ layers. This results in test likelihood similar to the that of the base model.

\section*{F. Masking Invalid Predictions}
As mentioned in Section \ref{subsec:vertex-model} we mask invalid predictions when evaluating our models. We identify a number of hard constraints that exist in the data, and mask the model's predictions that violate these constraints. The masked probability mass is uniformly distributed across the remaining valid values. We use the following masks:

\boldpar{Vertex model} 
\begin{itemize}
    \item The stopping token can only occur after an $x$-coordinate: 
    \begin{align} 
    v_k = s \implies v_k \text{ mod } 3 = 1
    \end{align}
    \item $z$-coordinates are non-decreasing:
    \begin{align}
        z_k \geq z_{k - 1}
    \end{align}
    \item $y$-coordinates are non-decreasing if their associated $z$-coordinates are equal: 
    \begin{align}
        y_k \geq y_{k - 1} \text{  if  } z_k = z_{k - 1}
    \end{align}
    \item $x$-coordinates are increasing if their associated $z$ and $y$-coordinates are equal:
    \begin{align}
        x_k > x_{k - 1} \text{  if  } y_k = y_{k - 1} \text{  and  } z_k = z_{k - 1}
    \end{align}
\end{itemize}

\boldpar{Face model}
\begin{itemize}
    \item New face tokens $n$ can not be repeated:
    \begin{align}
        f_k \neq n \quad \text{if} \quad f_{k - 1} = n
    \end{align}
    \item The first vertex index of a new face is not less than the first index in the previous face: 
    \begin{align} 
    f_1^{(k)} \geq f_{1}^{(k - 1)}, \quad k=1, \dotsc, N_f 
    \end{align}
    \item Vertex indices within a face are greater than the first index in that face:
    \begin{align}
        f_j^{(k)}  > f_1^{(k)}
    \end{align}
    \item Vertex indices within a face are unique: 
    \begin{align}
        f_i^{(k)}  \neq f_j^{(k)}, \quad \forall i, j
    \end{align}
    \item The first index of a new face is not greater than the lowest unreferenced vertex index:
    \begin{align}
        f_1^{(k)} \leq \text{min} \left[ \{v: v \leq N_V \} \setminus \{f_1^{(j)}, \dotsc, f_{N_j}^{(j)}\}_{j=1}^{k - 1} \right]
    \end{align}

\end{itemize}

\section*{G. Draco Compression Settings}
We compare our model in Table~\ref{table:log-likelihood} to Draco \cite{draco}, a performant 3D mesh compression library created by Google. We use the highest compression setting, quantize the positions to 8 bits, and do not quantize  in order to compare with the 8-bit mesh representations that our model operates on. Note that the quantization performed by Draco is not identical to our uniform quantization, so the reported scores are not directly comparable. Instead they serve as a ballpark estimate of the degree of compression obtained by existing methods.

\section*{H. Unconditional Samples}
\begin{figure*}[h]
    \centering
    \includegraphics[width=\linewidth]{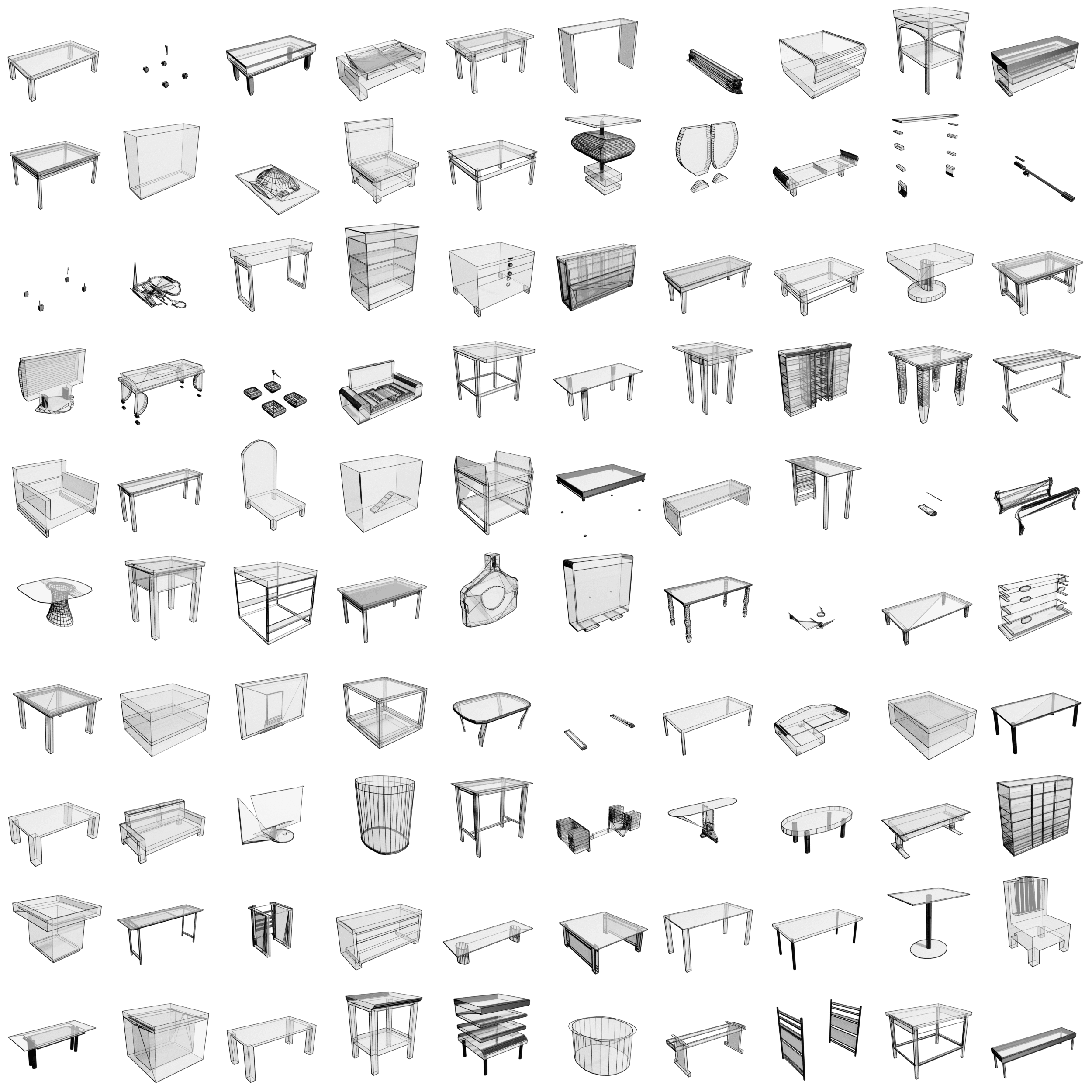}%
    \caption{Random unconditional samples using nucleus sampling with top-$p=0.9$.}
    \label{fig:unconditional}
\end{figure*}

Figure \ref{fig:unconditional} shows a random batch of unconditional samples generated using PolyGen with nucleus sampling ant top-$p=0.9$. The figure highlights . Firstly, the model learns to mostly output objects consistent with a shape class. Secondly, the samples contain a large proportion of certain object classes, including tables, chairs and sofas. This reflects the significant class-imbalance of the ShapeNet dataset, with many classes being underrepresented. Finally, certain failure modes are present in the collection. These include meshes with disconnected components, meshes that have produced the stopping token too early, producing incomplete objects, and meshes that don't have a distinct form that is recognizable as one of the shape classes.

\newcommand{\blocka}[2]{\multirow{3}{*}{\(\left[\begin{array}{c}\text{3$\times$3, #1}\\[-.1em] \text{3$\times$3, #1} \end{array}\right]\)$\times$#2}
}
\newcommand{\blockb}[3]{\multirow{3}{*}{\(\left[\begin{array}{c}\text{1$\times$1, #2}\\[-.1em] \text{3$\times$3, #2}\\[-.1em] \text{1$\times$1, #1}\end{array}\right]\)$\times$#3}
}
\newcommand{\vblocka}[2]{\multirow{3}{*}{\(\left[\begin{array}{c}\text{3$\times$3$\times$3, #1}\\[-.1em] \text{3$\times$3$\times$3, #1} \end{array}\right]\)$\times$#2}
}
\newcommand{\vblockb}[3]{\multirow{3}{*}{\(\left[\begin{array}{c}\text{1$\times$1$\times$1, #2}\\[-.1em] \text{3$\times$3$\times$3, #2}\\[-.1em] \text{1$\times$1, #1}\end{array}\right]\)$\times$#3}
}
\renewcommand\arraystretch{1.1}
\setlength{\tabcolsep}{3pt}
\begin{table*}[ht]
    \centering
    
    \begin{subtable}[b]{.475\textwidth}
    \centering
    \begin{tabular}{c|c|c}
    \hline
    layer name & output size & layer parameters \\
    \hline
    conv1 & 128$\times$128$\times$64 & 7$\times$7, 64, stride 2 \\
    \hline
    \multirow{4}{*}{conv2\_x} & \multirow{4}{*}{64$\times$64$\times$64} & 3$\times$3 max pool, stride 2 \\\cline{3-3}
      &  & \blocka{64}{1}  \\
      &  &  \\
      &  &  \\
    \hline
    \multirow{3}{*}{conv3\_x} &  \multirow{3}{*}{32$\times$32$\times$128}  & \blocka{128}{2}  \\
      &  & \\
      &  & \\
    \hline
    \multirow{3}{*}{conv4\_x} &  \multirow{3}{*}{16$\times$16$\times$256}  & \blocka{256}{2}  \\
      &  & \\
      &  & \\
    \hline
    & 256$\times$256  & spatial flatten (or) \\
    \hdashline
    & 1$\times$256  & average pool \\
    \hline
    \end{tabular}
    \caption{Image encoder}
    \end{subtable}
    \begin{subtable}[b]{.475\textwidth}
    \centering
    \begin{tabular}{c|c|c}
    \hline
    layer name & output size & layer parameters \\
    \hline
    embed & 28$\times$28$\times$28$\times$8 & embed, 8 \\
    \hline
    conv1 & 14$\times$14$\times$14$\times$64 & 7$\times$7$\times$7, 64, stride 2 \\
    \hline
    \multirow{3}{*}{conv2\_x} & \multirow{3}{*}{14$\times$14$\times$14$\times$64} & \vblocka{64}{1} \\
      &  &  \\
      &  &  \\
    \hline
    \multirow{3}{*}{conv3\_x} &  \multirow{3}{*}{7$\times$7$\times$7$\times$256}  & \vblocka{256}{2}  \\
      &  & \\
      &  & \\
    \hline
    & 343$\times$256  & spatial flatten (or) \\
    \hdashline
    & 1$\times$256  & average pool \\
    \hline
    \end{tabular}
    \caption{Voxel encoder}
    \end{subtable}
    \caption{Architectures for image and voxel encoders. Pre-activation residual blocks are shown in brackets, with the numbers of blocks stacked. Downsampling is performed by conv3{\_}1, conv4{\_} for image encoders, and by conv3{\_} for voxel encoders, with a stride of 2. For AtlasNet models, we perform an additional linear projection up to 1024 dimensions before average pooling to obtain a vector shape representation.
    }
    \label{tab:arch}
\end{table*}

\end{document}